\documentclass[aps,pre,onecolumn,superscriptaddress,showpacs]{revtex4-1}
\usepackage{graphicx}
\usepackage{dcolumn}
\usepackage{bm}

\usepackage{natbib}
\usepackage[
text={7in,9.5in},centering,
]{geometry}

\usepackage{float}
\usepackage{epsfig}
\usepackage{amsmath}
\usepackage{amssymb}
\usepackage{textcomp}
\usepackage{layouts}

\begin{document}

\title{Proper Time and Mass of the Universe}

\author{Natalia Gorobey}
\affiliation{Peter the Great Saint Petersburg Polytechnic University, Polytekhnicheskaya
29, 195251, St. Petersburg, Russia}

\author{Alexander Lukyanenko}\email{alex.lukyan@mail.ru}
\affiliation{Peter the Great Saint Petersburg Polytechnic University, Polytekhnicheskaya
29, 195251, St. Petersburg, Russia}

\author{A. V. Goltsev}
\affiliation{Ioffe Physical- Technical Institute, Polytekhnicheskaya
26, 195251, St. Petersburg, Russia}

\begin{abstract}
A modification of the homogeneous isotropic model of the Friedman universe with a scalar field is proposed, in which the proper time of the universe is added to the dynamic variables under the additional condition of its classical dynamics. Along with proper time, an additional parameter is included in the modified theory, which is an analog of the rest mass of a relativistic particle. Assuming the dependence of the gravitational constant on proper time, we obtain the rest mass as an additional dynamic variable.
\end{abstract}


\maketitle

\section{Introduction}

In covariant theories, the symmetry group of which includes the reparametric invariance of time, proper time naturally arises \cite{1,2}. The simplest and most illustrative example here is a relativistic particle with action in a parameterized form:
\begin{equation}
 I=\frac{1}{2} \int_0^1 d\tau(\frac{\dot{x}^2}{N}+ m^2 N),
 \label{1}
\end{equation}
where $\tau$ is an arbitrary parameter on the particle's world line, and $N(\tau)$ is an additional variable (lapse function in General Relativity \cite{3}). We use the abbreviation for the square of 4-speed in Minkowski space. Proper time
\begin{equation}
 s=\int_0^1 d\tau N(\tau)
 \label{2}
\end{equation}
is invariant with respect to infinitesimal reparametrization
\begin{equation}
 \delta{N}=\dot{\epsilon}
  \label{3}
\end{equation}
if the boundary conditions are satisfied
\begin{equation}
 \epsilon(0)=\epsilon(1)=0.
  \label{4}
\end{equation}
Reparametrization includes dynamics, the generator of which is the Hamiltonian constraint quadratic in momenta ($c=1$),
\begin{equation}
H=p^2-m^2.
  \label{5}
\end{equation}
The same situation with time in the dynamics of a closed universe in the framework of the theory of gravity \cite{3}. Proper time is not observable in quantum theory, where additional averaging is provided for it. Such is the structure of the covariant quantum theory in the approach based on BRST invariance \cite{4,5,6,7,8}. In the case of a relativistic particle, the BRST-invariant functional integral includes additional integration over the proper time \cite{9}. In \cite{10}, it was proposed to include the proper time in a set of dynamic variables. In the case of a relativistic particle, this is achieved by the following replacement, 
\begin{equation}
N=\dot{s}
 \label{6}
\end{equation}
($\delta{s}=\epsilon$). In order for this variable to make sense of the evolution parameter, in \cite{11} an additional condition was imposed on it, preserving the classical dynamics in quantum theory.
In this paper, this two-step modification is proposed in the dynamics of a homogeneous model of the Friedman universe with a scalar field of matter. Together with proper time, a canonical momentum conjugate to its infinitesimal variation enters into the modified theory. In relativistic mechanics, this momentum is proportional to (squared) the rest mass of the particle \cite{11}. 
By analogy, we will call it the rest mass of the universe. This value is an integral of motion.
The introduction of proper time into the dynamics of the universe as an evolution parameter allows us to bring into discussion the hypothesis of variation of the gravitational constant $G$. If we consider $G$ as a function of proper time, the rest mass of the universe is not an integral of motion.

\section{Modification of the Friedman model}
The original Lagrangian of a homogeneous isotropic model of the Friedman universe is
\begin{equation}
 L=\frac{1}{2g}(\frac{a\dot{a}^2}{N}-aN)-2\pi^2a^3\frac{1}{2}(\frac{\dot{\phi}^2}{N}-V(\phi)N).
\end{equation}
As in the case of a relativistic particle, it contains an arbitrary function of time $N(t)$ (lapse function \cite{3}). Here, we consider a single scalar field as matter $\phi$. We will carry out both stages of the modification of this model immediately. First, we introduce the proper time, replacing the lapse function according to (\ref{6}), and then we make the final shift of the proper time by a value $\epsilon$. As a result, we get:
\begin{eqnarray}
 \tilde{L}=\frac{1}{2g(s)}[\frac{a\dot{a}^2}{\dot{s}}(1+\frac{g'\epsilon}{g}+\frac{\dot{\epsilon}}{\dot{s}})-a(\dot{s}(1-\frac{g'\epsilon}{g})-\dot{\epsilon})]
 \notag \\
-2\pi^{2} a^3 \frac{1}{2} [\frac{\dot{\phi}^2}{\dot{s}}(1+\frac{\dot{\epsilon}}{\dot{s}})-V(\dot{s}-\dot{\epsilon})].
 \label{8}
\end{eqnarray}
Here, we  admit the possibility of the dependence of the gravitational constant on proper time, although in this paper we will consider it as a constant. The effect of the variation of the gravitational constant on the dynamics of the universe will be discussed at the end. We turn to the canonical form of the theory. We find the canonical momenta for all dynamic variables:
\begin{equation}
 p_a=\frac{1}{g}\frac{a\dot{a}}{\dot{s}}(1+\frac{\dot{\epsilon}}{\dot{s}}),
\end{equation}
\begin{equation}
 p_{\phi}=-2{\pi}^2a^3\frac{\dot{\phi}}{\dot{s}}(1+\frac{\dot{\epsilon}}{\dot{s}}),
\end{equation}
\begin{equation}
 p_s=-\frac{1}{2g}[\frac{a\dot{a}^2}{\dot{s}^2}(1+2\frac{\dot{\epsilon}}{\dot{s}})+a]+2\pi^2a^3\frac{1}{2}[\frac{\dot{\phi}^2}{\dot{s}^2}(1+2\frac{\dot{\epsilon}}{\dot{s}})+V],
\end{equation}
\begin{equation}
 P_{\epsilon}=\frac{1}{2g}(\frac{a\dot{a}^2}{\dot{s}^2}+a)-2\pi^2a^3\frac{1}{2}(\frac{\dot{\phi}^2}{\dot{s}^2}+V).
 \label{12}
\end{equation}
The Hamiltonian is equal to zero, as it should be for the reparameterization-invariant theory. But we have a Hamiltonian constraint, which is now not quadratic in momenta:
\begin{equation}
 p_s=P_{\epsilon}-\frac{a}{g}+a^3V-2\sqrt{P_{\epsilon}-\frac{a}{2g}+\frac{a^3}{2}V}\sqrt{\frac{g}{2a}p_a^2-\frac{1}{2a^3}p_{\phi}^2}.
 \label{13}
 \end{equation}
Here the square roots appear as a result of the exclusion of the generalized velocity, according to
\begin{equation}
 \frac{\dot{\epsilon}}{\dot{s}}=-1+\frac{\sqrt{\frac{g}{2a}p_a^2-\frac{1}{2a^3}p_{\phi}^2}}{\sqrt{P_{\epsilon}-\frac{a}{2g}+\frac{a^3}{2}V}}.
\end{equation}
Their appearance reflects the classical character of the dynamics of proper time.
According to (\ref{12}), the value $P_{\epsilon}$ is the resulting energy of the universe, which before modification was equal to zero. This energy has nothing to do with the physical degrees of freedom, but is entirely related to the proper time of the universe. Being a constant, it is specified as a parameter of the model, along with other constants included in the original Lagrangian. This value should not be confused with the cosmological term, which is proportional to $a^3$. By analogy with \cite{11}, this quantity can be called the mass of the universe. 

If we allow the variation of the gravitational constant ($g'\neq0$), the structure of the Hamiltonian constraint will change, so that $P_{\epsilon}$ becomes a dynamic variable, equivalent with fundamental degrees of freedom. A change in the state of the universe with the passage of its proper time is accompanied by a change its rest mass. 
If we exclude the dependence of the gravitational constant on proper time, we can also refuse to introduce the rest mass and put $P_{\epsilon}=0$. Then, taking $s=\tau$, the Hamiltonian constraint may be written as follows:
\begin{equation}
\sqrt{\frac{g}{2a}p_a^2-\frac{1}{2a^3}p_{\phi}^2}=\sqrt{-\frac{a}{2g}+\frac{a^3}{2}V}
\end{equation}

It is easy to see that as a result we obtain the square root of the Hamiltonian constraint of the original theory. In quantum theory, the square root of the quadratic form of momenta can be extracted by considering the wave function of the universe as a spinor. In this case, two Dirac matrices \cite{12} are enough for us to extract the root, and we get an analogue of the Dirac equation for the Friedman universe:
\begin{equation}
i\hslash[\hat{\gamma}_0\sqrt{\frac{g}{2a}}\frac{\partial}{\partial{a}}+\hat{\gamma}_1\sqrt{\frac{1}{2a^3}}\frac{\partial}{\partial{\phi}}]\psi=\sqrt{-\frac{a}{2g}+\frac{a^3}{2}V}\psi.
\end{equation}
Now the role of proper time can be played by any physical dynamic variable. An additional observable - the “spin” of the universe should be understood then as a parameter of the reference frame in which proper time is measured.
\section{Conclusions}
Thus, the introduction of proper time as a classical dynamic variable in cosmology is accompanied by a corresponding contribution to the energy balance of the universe. Considering this additional energy to be observable, we obtain a method for measuring proper time. The proposed formalism also allows us to interpret the alleged variation of the gravitational constant as the dependence of the latter on proper time. In the presence of such a dependence, the energy of proper time becomes a dynamic variable added to the initial fundamental set.
The modification of the Friedman model considered here illustrates the general (two-stage) approach to quantization of covariant theories, formulated at the beginning. In this case, we have a reparametrization-invariant theory with one Hamiltonian constraint. In the algebra of constraints of the theory of gravitation, we have a set of Hamiltonian constraints that generate multi-arrow proper time, as well as linear in momentum constraints, generating spatial translations. In the modified theory of gravity, one should expect the appearance of the corresponding local density of rest mass of the universe, as well as its flows in space.
\section{acknowledgements}
The authors thank V.A. Franke for useful discussions.

\end{document}